\documentclass[11pt,a4paper]{article}
\usepackage{a4wide}
\usepackage{hyperref}
\usepackage{float}
\usepackage{bm}
\usepackage{xcolor}
\usepackage{graphicx}
\usepackage{etoolbox}
\usepackage{subcaption}
\usepackage{authblk}
\usepackage{mathcomp}

\usepackage{amsmath} 
\usepackage{amsthm} 
\usepackage{hyperref} 
\usepackage{graphics} 
\usepackage{algorithmic} 
\usepackage{url} 

\begin{document}

\begin{center}
{\Large
A novel method to determine the 4th moment of the $^{208}$Pb charge density from low-momentum-transfer electron scattering
}

\vspace{5mm} 
\noindent
  
Rika Danjo$^1$, Toshimi Suda$^{1,2}$ and Toshio Suzuki$^1$
 
$^1$Research Center for Accelerator and Radioisotope Science,
Tohoku University, Sendai 982-0826, Japan\\
 $^2$Department of physics, Tohoku University, Sendai 980-8578, Japan\\

\end{center}

\vspace{3mm}
\noindent
e-mail:rika.danjo.t8@dc.tohoku.ac.jp

\vspace{5mm}
\noindent
Abstract :
We propose a new method to determine the 4th moment of the nuclear charge-density distribution of $^{208}$Pb, $\langle r^{4}\rangle_{c}$, with high precision from low-momentum-transfer elastic electron scattering. 
The key feature of this approach is that Coulomb distortion effects, which are pronounced for heavy nuclei, are evaluated numerically and incorporated into the formalism of the Plane Wave Born Approximation as a correction factor. 
The 4th and higher-order charge-density moments are of particular interest because they are sensitive not only to the moments of the point-proton density distribution but also to those of the point-neutron density distribution in nuclei. Their precise determination is therefore essential for constraining the moments of the point-neutron density distribution. 
It is demonstrated that $\langle r^{4}\rangle_{c}$ can be determined with an uncertainty of about 0.8$\%$ using the proposed approach, improving upon the present uncertainty of about 1.5$\%$ obtained from empirically determined charge-density distributions.
Measurements of $\langle r^{4}\rangle_{c}$ in $^{208}$Pb using this method are currently in progress at the Ultra-Low momentum transfer Q$^{2}$ (ULQ$^{2}$) facility at Research Center for Accelerator and Radioisotope Science (RARiS), Tohoku University. 
\vspace{1cm}

\section{Introduction}
    The charge density distributions of nuclei, $\rho_{c}$($r$), are best determined by elastic electron scattering \cite{eN}, and those of the stable nuclei studied thus far have played an essential role in the understanding of their internal structure \cite{deVries_1987}. 
    
    In addition to $\rho_{c}(r)$ itself, the $n$th moment of $\rho_{c}$($r$) is also an important quantity, defined as
    \begin{equation}
        \label{eq:def_moment}
        \langle r^{n} \rangle_{c} \equiv \frac{1}{Z} \int r^{n}\rho_{c}({r})\,\mathrm{d}^{3}{r},
    \end{equation}  
    where $Z$ denotes the atomic number.
    The 2nd moment of $\rho_{c}$($r$), $\langle r^2 \rangle_{c}$, represents the mean-square radius (msr), which has been precisely determined for both stable and unstable nuclei through the combined analysis of elastic electron scattering, muonic X-ray, and isotope shifts \cite{r2_accu}. 
    The msr is expressed as \cite{suzuki_2019, suzuki_2021}
    \begin{equation}
	\label{eq:2th_Rn}
	\langle r^2 \rangle_{c} = \langle r^2\rangle_{p} +r^2_{p} + \frac{N}{Z}r^2_{n} + {\Delta{w_{2,\,\tau}}},
    \end{equation}
     where $\langle r^{2}\rangle_{p}$, $r^{2}_{\tau}$, $N$, and $\Delta{w_{2,\,\tau}}$ ($\tau$ = $n$, $p$) represent the msr of the point-proton density distribution, the msr of the nucleon charge density distribution, the number of neutrons in the nucleus, and the msr of the spin-orbit density distribution, respectively. Hereafter, we denote $\langle r^{n}\rangle_{\tau}$ as the $n$th moment of the point-nucleon density distribution. 
    
    Recently, the 4th moment of $\rho_{c}$($r$), $\langle r^4 \rangle_{c}$, has been shown to have a relation to the msr of the point-neutron density, $\langle r^2 \rangle_{n}$ \cite{suzuki_2019, suzuki_2021};
    \begin{equation}
	\label{eq:4th_Rn}
	\langle r^4 \rangle_{c} = \langle r^4\rangle_{p} +\frac{10}{3} \langle r^2 \rangle_{p} r^2_{p} + \frac{N}{Z} \frac{10}{3} \langle r^2 \rangle_{n} r^2_{n} + {\Delta_{W_{\tau}}} + {\Delta_{4}},
    \end{equation}
    where $\Delta_{W_{\tau}}$ and $\Delta_{4}$ represent the contribution from the moment of the spin-orbit density and the 4th moment of the nucleon charge density, respectively.
   The precise value of $\langle r^2 \rangle_{n}$ remains under discussion \cite{deltaR}. In this context, Eq. \eqref{eq:4th_Rn} provides a new way to access $\langle r^2 \rangle_{n}$ through elastic electron scattering, based solely on the electromagnetic interaction.

   To date, the only practical way to determine $\langle r^{4}\rangle_{c}$ has been to extract $\rho_c(r)$ from elastic electron scattering and apply the extracted $\rho_c(r)$ to Eq.~\eqref{eq:def_moment}.
   In this approach, scattering cross sections need to be measured over a wide range of momentum transfer, $q$. Here, $q$ denotes the momentum transferred from the incident electron to the nucleus. The measured cross sections are analyzed using "model-independent" methods, such as the sum of Gaussians (SOG) analysis \cite{SOG} and the Fourier-Bessel (FB) analysis \cite{FB}, to extract $\rho_{c}(r)$.
   
   Although the FB analysis has provided $\rho_{c}(r)$ for many stable nuclei, the determined $\rho_{c}(r)$ for ${}^{208}$Pb is the most precise among them \cite{ePb}. This is largely because elastic electron scattering data for $^{208}$Pb cover the widest $q$ range, $0.44\,\mathrm{fm}^{-1} \leq q \leq 3.70\,\mathrm{fm}^{-1}$, compared with those for other stable nuclei \cite{deVries_1987}. Within this framework, the uncertainty of $\langle r^{4}\rangle_{c}$ for $^{208}$Pb is about $1.5\%$ \cite{deVries_1987, suzuki_2021, ePb}.
It should be noted that since correlations among the FB coefficients are not taken into account, this uncertainty may not be accurately estimated. 

The neutron contribution to $\langle r^{4}\rangle_{c}$ in $^{208}\mathrm{Pb}$ has been theoretically estimated to be only about $2$--$3\%$ \cite{suzuki_2021}. Since $\langle r^{4}\rangle_{c}$ has so far been determined with a precision of approximately $1.5\%$ \cite{deVries_1987, suzuki_2021, ePb}, this precision already provides meaningful sensitivity to such a small contribution. Indeed, this sensitivity has been used to constrain the root-mean-square radius of the point-neutron density distribution, yielding $\sqrt{\langle r^2\rangle_n}=5.728(57)\,\mathrm{fm}$ under the relativistic frameworks and $\sqrt{\langle r^2\rangle_n}=5.609(54)\,\mathrm{fm}$ under the non-relativistic frameworks \cite{suzuki_2021}. A further improvement in the experimental precision of $\langle r^{4}\rangle_{c}$ would therefore enable a more precise extraction of $\sqrt{\langle r^2\rangle_n}$ within these theoretical frameworks.
   
In this paper, we propose a new method to determine $\langle r^4 \rangle_c$
with the aim of improving its precision compared with previous studies  \cite{deVries_1987, suzuki_2021, ePb}.
The advantage of the present method is that it requires only additional elastic-scattering data in the low-$q$ range,
$q < 0.3~\mathrm{fm}^{-1}$, together with the data already available,
thereby avoiding the further difficult measurements in the high-$q$ range. It should be noted that the scattering cross section is proportional to $1/q^4$.
   
\section{Form factor in PWBA and the $n\mathrm{th}$ moment}

    Elastic electron scattering cross section, $\left(\frac{\mathrm{d}\sigma}{\mathrm{d}\Omega}\right)_\mathrm{PWBA}$, under the Plane Wave Born Approximation (PWBA) for spin 0$^{+}$ nuclei is given by
    \begin{equation}
    \label{eq:PWBA}
    \left(\frac{\mathrm{d}\sigma}{\mathrm{d}\Omega}\right)_\mathrm{PWBA}=\left(\frac{\mathrm{d}\sigma}{\mathrm{d}\Omega}\right)_{\mathrm{Mott}}{|F_\mathrm{PWBA}({q})|^2},
    \end{equation}
    where
$\left(\frac{\mathrm{d}\sigma}{\mathrm{d}\Omega}\right)_{\mathrm{Mott}}$
is the Mott cross section for elastic scattering from a point-like charged nucleus:
\begin{equation}
\label{eq:def_mott}
\left( \frac{\mathrm{d}\sigma}{\mathrm{d}\Omega} \right)_{\mathrm{Mott}}
=
\frac{\alpha^2 \cos^2(\theta/2)}{4E^2 \sin^4(\theta/2)}
\cdot
\frac{E'}{E}.
\end{equation}
Here, $\alpha$, $E$, $E'$, and $\theta$ denote the fine-structure constant, the incident electron energy, the scattered electron energy, and the scattering angle, respectively.
The quantity $F_\mathrm{PWBA}(q)$ is the charge form factor. When expressed in terms of $\rho_c(r)$, it is given by

    \begin{equation}
        \label{eq:def_rho}
       F^\mathrm{\rho_c(r)}_\mathrm{PWBA}(q) = \frac{1}{Z}\int \rho_c(r)\,\mathrm{e}^{i\bm{q} \cdot \bm{r}}\,{\mathrm{d}}^{3}{r}.
    \end{equation}
    The normalization condition is given by $\int \rho_{c}(r)\,\mathrm{d}^{3}{r}$ = $Z$.
     
\begin{figure*}[t]
  \centering
    \centering
    \includegraphics[width=\linewidth]{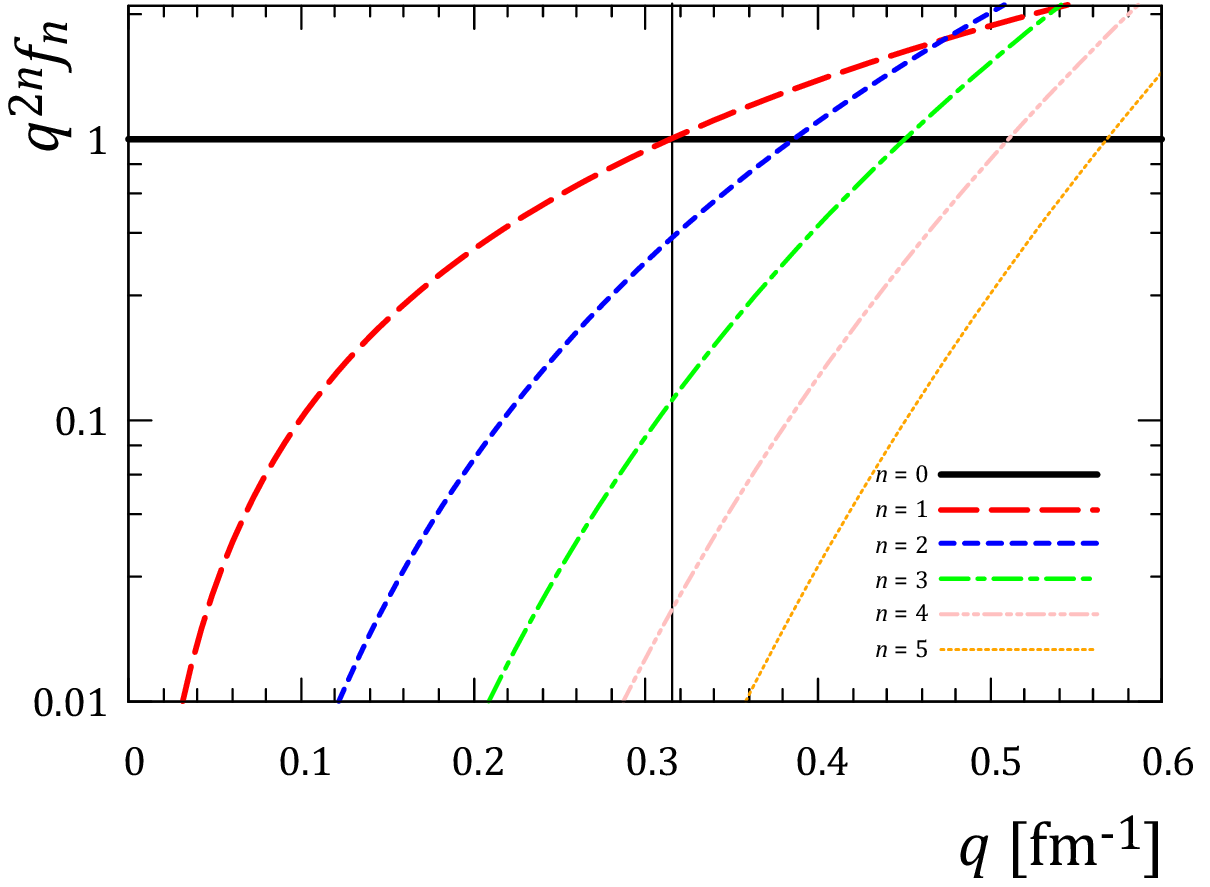}
    \label{fig:fk_Pb}
  \hfill
   \caption{Behavior of $q^{2n}f_{n}$ as a function of $q$ for $^{208}$Pb. 
  Each line represents the $n{\mathrm{th}}$-order term of Eq. \eqref{eq:Taylor_moment}. The vertical black line shows the convergence condition for $q < 0.315~\mathrm{fm}^{-1}$. }
  \label{fig:fk_all}
\end{figure*}

    The Taylor expansion of $|F_\mathrm{PWBA}(q)|^2$ in terms of $\langle r^{n}\rangle_{c}$ is shown as
     \begin{equation}
    \begin{aligned}
    \label{eq:Taylor_moment}
       |F_\mathrm{PWBA}(q)|^2
       &=  \sum_{n=0}^\infty(-1)^{n} q^{2n} f_{n}\\
       &= \sum_{n=0}^\infty(-1)^{n} q^{2n} \sum_{k=0}^n \frac{\langle r^{2k}\rangle_{c} \langle r^{2(n-k)}\rangle_{c}}{(2k+1)!(2n-2k+1)!}\\
       &= 1 -\frac{\langle r^2 \rangle_c}{3}q^2 + \frac{(3\langle r^4 \rangle_c + 5\langle r^2 \rangle_c^2)}{180}q^4 -\frac{(\langle r^6 \rangle_c + 7\langle r^4 \rangle_c \langle r^2 \rangle_c)}{2520}q^6 + \cdots.\\ 
    \end{aligned}
    \end{equation} 
To use the truncated Taylor expansion, it is necessary to examine the range of $q$ over which the magnitudes of successive terms in Eq. \eqref{eq:Taylor_moment} decrease monotonically, as required by the Leibniz criterion for convergent of an alternating series.
This condition can be written as
\begin{equation}
q^{2n}f_n > q^{2(n+1)}f_{n+1}\quad (n=0,1,2,\cdots).
\end{equation}
Explicitly, the first few terms should satisfy
\begin{equation}
\begin{aligned}
1 > \frac{\langle r^{2} \rangle_{c}}{3}\, q^{2} > \frac{(3\langle r^4 \rangle_c + 5\langle r^2 \rangle_c^2)}{180}q^4 > \frac{(\langle r^6 \rangle_c + 7\langle r^4 \rangle_c \langle r^2 \rangle_c)}{2520}q^6 >\cdots.
\end{aligned}
\end{equation}
Although a rigorous proof of the monotonic decrease for arbitrary $n$ is not given here, the behavior of the terms $q^{2n}f_n$ shown in Fig.~\ref{fig:fk_all} for $^{208}$Pb indicates that the condition imposed by the first two terms is the most restrictive. Therefore, we use this condition as a practical criterion:
\begin{equation}
1 > \frac{\langle r^{2} \rangle_{c}}{3} q^{2}.
\end{equation}
Using the value of $\langle r^2 \rangle_{c}$ from Ref.~\cite{r2_accu}, this criterion gives the upper limit of $q < 0.315~\mathrm{fm}^{-1}$ for $^{208}$Pb. 

As shown in Fig.~\ref{fig:fk_all}, $q^{2n}f_n$ in Eq.~\eqref{eq:Taylor_moment} decreases rapidly with increasing $n$, so that $|F_\mathrm{PWBA}(q)|^2$ can be approximated by truncating the series at a finite order. However, careful judgment is required in deciding up to which order terms in the series should be included.
This is because the accuracy of the approximation is governed by the magnitude of the neglected higher-order terms. More specifically, when the expansion is truncated at order $q^{2k}$, $|F_\mathrm{PWBA}(q)|^2$ can be expressed as
\begin{equation}
|F_\mathrm{PWBA}(q)|^2 \simeq \sum_{n=0}^{k} (-1)^n q^{2n} f_n.
\end{equation}
The accuracy of the approximation is governed by the size of the omitted higher-order terms, $q^{2n}f_n$ with $n \geq k+1$.
Their effects can be assessed by examining whether the extracted $\langle r^{4}\rangle_c$ over the $q$ range of interest stabilizes, as the highest order retained in the expansion of Eq.~\eqref{eq:Taylor_moment} is increased. These points are discussed in detail in Sect.~4.

Thus, within the PWBA framework, the value of \(\langle r^{4}\rangle_c\) of $^{208}$Pb can be extracted from the Taylor expansion of \(|F_\mathrm{PWBA}(q)|^2\) using elastic electron scattering data in the range of \(q < 0.315~\mathrm{fm}^{-1}\).
 
\section{Coulomb correction}

In actual electron scattering, the electron waves are seriously distorted by the Coulomb field of the nucleus. This distortion limits the applicability of the PWBA expression, even for light nuclei, and its effect becomes more significant for heavy nuclei such as $^{208}$Pb.
Therefore, the effects of Coulomb distortion must be corrected to obtain $|F_\mathrm{PWBA}(q)|^2$ that can be used for the extraction of $\langle r^{4}\rangle_c$.

Corrections due to Coulomb distortion effects have been investigated by many authors \cite{SOG, Tamae, Amroun, qeff}. A commonly used way to retain the PWBA framework is to employ the effective momentum approximation (EMA), in which $q$ in Eq. \eqref{eq:PWBA} is replaced by the effective momentum transfer, $q_{\mathrm{eff}}$ \cite{qeff, qeff_exp}. The effective momentum transfer is introduced to account for the change in the electron momentum caused by the nuclear Coulomb field during scattering and is commonly defined as
$q_{\mathrm{eff}} = q\left(1-\frac{V_c}{E}\right)$,
where $V_c$ is the effective Coulomb potential energy experienced by the electron and is determined phenomenologically. 

\begin{figure*}[t]
  \centering
    \centering
    \includegraphics[width=\linewidth]{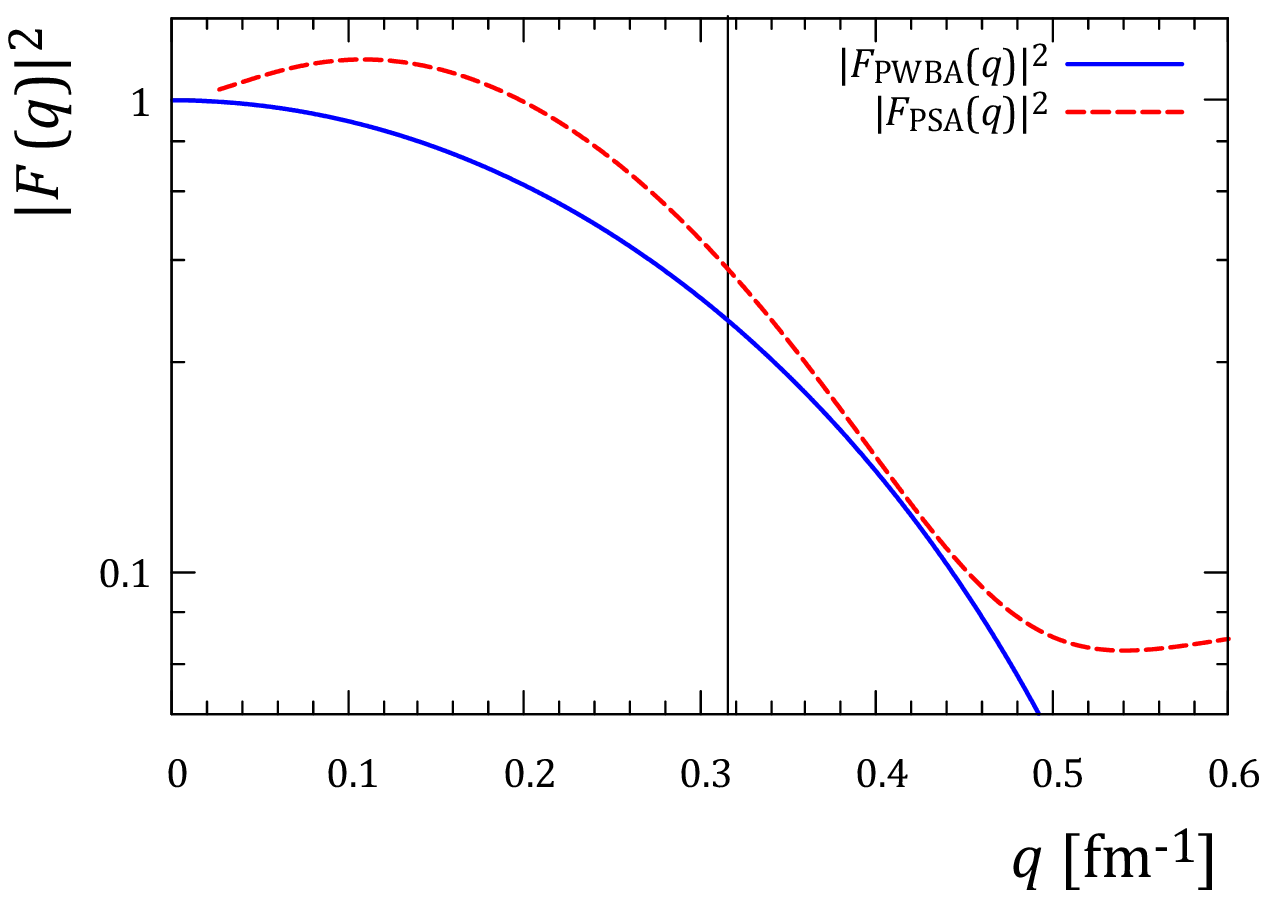}
    \label{fig:Coulomb_effect_Pb}
  \caption{Coulomb distortion effects calculated using the FB  distribution for $^{208}$Pb \cite{deVries_1987}. The red and blue lines represent $|F_\mathrm{PWBA}^{\mathrm{FB}}(q)|^2$ and $|F^{\mathrm{FB}}_\mathrm{PSA}(q)|^2$, respectively. The difference between two lines is the effects of Coulomb distortion. The vertical black line shows the convergence condition for $q$. For details, see the text.}
  \label{fig:Coulomb_effect}
\end{figure*}

The validity of the EMA in the low-$q$ range relevant to the present study has not been examined so far.
To investigate Coulomb distortion effects for \(^{208}\)Pb in the low-$q$ range, we compare the cross sections calculated within the phase-shift analysis (PSA) \cite{DREPHA} using the FB distribution \cite{deVries_1987} with those calculated within the PWBA framework using the same distribution.
Figure \ref{fig:Coulomb_effect} shows the Coulomb distortion effects expressed in terms of the form factor.
The blue line shows the squared form factor obtained in the PWBA, $|F^{\mathrm{FB}}_\mathrm{PWBA}(q)|^2$ = $\left(\frac{\mathrm{d}\sigma}{\mathrm{d}\Omega}\right)^{\mathrm{FB}}_\mathrm{PWBA}$/$\left(\frac{\mathrm{d}\sigma}{\mathrm{d}\Omega}\right)_\mathrm{Mott}$, and the red one the squared form factor defined as $|F^{\mathrm{FB}}_\mathrm{PSA}(q)|^2$ = $\left(\frac{\mathrm{d}\sigma}{\mathrm{d}\Omega}\right)^{\mathrm{FB}}_\mathrm{PSA}$/$\left(\frac{\mathrm{d}\sigma}{\mathrm{d}\Omega}\right)_\mathrm{Mott}$. Here $\left(\frac{\mathrm{d}\sigma}{\mathrm{d}\Omega}\right)^{\mathrm{FB}}_{\mathrm{PSA}}$ is calculated by the PSA \cite{DREPHA} at $E = 60$ MeV, with $\theta = 5^\circ$--$175^\circ$.
Note that $|F^{\mathrm{FB}}_\mathrm{PWBA}(q)|^2$ and  $|F^{\mathrm{FB}}_\mathrm{PSA}(q)|^2$ are plotted as a function of $q$, which is given by
\begin{equation}
    \label{eq:q_ela}
   q = 2E\,\sin(\frac{\theta}{2}).
\end{equation}
The observed difference between the blue and red lines represents the effect of Coulomb distortion. In the low-$q$ range, the result of $|F^{\mathrm{FB}}_\mathrm{PSA}(q)|^2$ shows a bump-like structure that is absent in the result of $|F^{\mathrm{FB}}_\mathrm{PWBA}(q)|^2$. Such a structure cannot be reproduced by the EMA, which only replaces $q$ with $q_{\mathrm{eff}}$ in $|F_\mathrm{PWBA}^\mathrm{FB}(q)|^2$.

\begin{figure*}[t]
  \centering
    \centering
    \includegraphics[width=\linewidth]{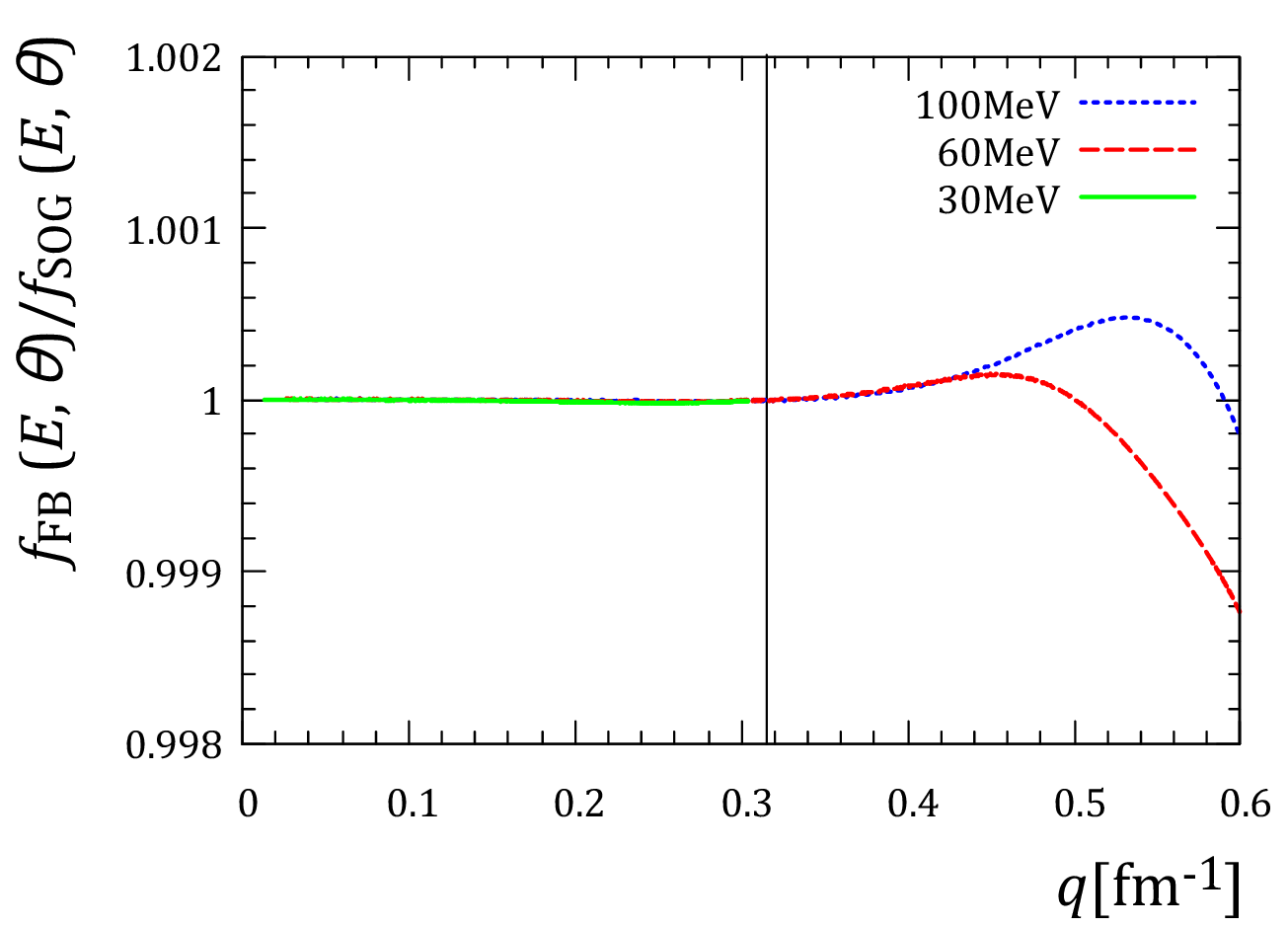}
    \label{fig:Coulomb_comp_Pb}
  \caption{The ratio $f_\mathrm{FB}(E,\theta)/f_\mathrm{SOG}(E,\theta)$ of the Coulomb distortion correction factors for $^{208}$Pb, where $f_\mathrm{FB}(E,\theta)$ and $f_\mathrm{SOG}(E,\theta)$ are calculated using the FB and SOG distributions, respectively. They are calculated for $E$ = 30, 60, and 100 MeV and over $\theta$ = 5$\tcdegree$ - 175$\tcdegree$, corresponding to momentum-transfer ranges of $q=0.013\text{--}0.304\ \mathrm{fm}^{-1}$, $0.027\text{--}0.670\ \mathrm{fm}^{-1}$, and $0.044\text{--}1.014\ \mathrm{fm}^{-1}$, respectively. The vertical black line shows the convergence condition for $q$.}
  \label{fig:Coulomb_comp}
\end{figure*} 

To describe the Coulomb distortion effects observed in Fig. \ref{fig:Coulomb_effect}, we treat them as a multiplicative correction to the PWBA expression, inspired by a similar treatment of three-body systems presented in Ref. \cite{3He-3H}. The scattering cross section for elastic electron scattering is assumed to be expressed as
\begin{equation}
    \label{eq:PWBA_corr}
    \left(\frac{\mathrm{d}\sigma(E,\theta)}{\mathrm{d}\Omega}\right)^{\rho(r)}_\mathrm{PSA}=f_{\rho_{c(r)}}(E,\theta)\left(\frac{\mathrm{d}\sigma(E,\theta)}{\mathrm{d}\Omega}\right)_{\mathrm{Mott}}{|F_\mathrm{PWBA}^\mathrm{\rho(r)}({q})|^2}.
\end{equation}

Since $f_{\rho_{c}(r)}(E,\theta)$ depends on both $\rho_c(r)$ and $E$, 
when $\theta$ is fixed, we examine their dependence explicitly.
Two "model-independent" $\rho_c(r)$, namely the FB and SOG distributions, are used for, with three incident energies, $E = 30$, $60$, and $100\,\mathrm{MeV}$, which satisfy the convergence condition over a wide range of \(q\).
For each combination of $\rho_c(r)$ and $E$, the scattering cross section,  $\left(\frac{\mathrm{d}\sigma(E,\theta)}{\mathrm{d}\Omega}\right)_{\mathrm{PSA}}^{\rho_c(r)}$, is calculated using PSA \cite{DREPHA} over the range of $\theta = 5^\circ$--$175^\circ$, and $|F(q)_{\mathrm{PSA}}^{\rho_c(r)}|^2$ is obtained from Eq.~\eqref{eq:def_rho}. The quantity \(f_{\rho_c(r)}(E,\theta)\) is then derived from $\left(\frac{\mathrm{d}\sigma(E,\theta)}{\mathrm{d}\Omega}\right)_{\mathrm{PSA}}^{\rho_c(r)}$
and
$|F(q)_{\mathrm{PSA}}^{\rho_c(r)}|^2$
using Eq.~\eqref{eq:PWBA_corr}, as shown in Fig.~\ref{fig:Coulomb_comp}.

In the low-$q$ range satisfying the convergence condition, it is found that the relative difference between $f_{\mathrm{FB}}(E,\theta)$ and $f_{\mathrm{SOG}}(E,\theta)$ satisfies
$\frac{\left|f_{\mathrm{FB}}(E,\theta)-f_{\mathrm{SOG}}(E,\theta)\right|}{f_{\mathrm{FB}}(E,\theta)} < 10^{-4}$ for all $E$ considered. This indicates that the correction factor is insensitive to the FB and SOG charge density distributions. Moreover, the energy dependence is found to be small.
Therefore, within this low-\(q\) range, neither the choice of $E$ nor the choice between the FB and SOG charge density distributions imposes a  constraint on the extraction of \(\left|F_\mathrm{PWBA}(q)\right|^2\).

\section{Experimental determination of $\langle r^4 \rangle_{c}$ of $^{208}$Pb}

In this section, we describe the procedure for determining \(\langle r^{4} \rangle_{c}\) of \(^{208}\mathrm{Pb}\) based on the extracted \(|F_\mathrm{PWBA}(q)|^{2}\), and evaluate the uncertainty in $\langle r^{4} \rangle_{c}$ obtained with the proposed method.
 
The precision of the extracted \(\langle r^{4} \rangle_{c}\) depends on the $q$ range included in the analysis, whereas its accuracy depends on the order at which the expansion of Eq.~\eqref{eq:Taylor_moment} is truncated. Hereafter, these are referred to as conditions (i) and (ii), respectively.
The condition (i) is chosen to satisfy the convergence of the series and maximize the precision in the determination of $\langle r^{4} \rangle_{c}$.
Once the $q$ range is fixed, the order in condition (ii) is determined so that the truncation error arising from omitted higher-order terms in the expansion of Eq. \eqref{eq:Taylor_moment} is negligible. This requirement is confirmed when the extracted value of $\langle r^{4} \rangle_{c}$ becomes independent of $q$, as discussed later.

  \begin{table}
     \begin{center}
     \caption{The \(n\)th moments of the FB, SOG, and 2pF density distributions of \(^{208}\)Pb, in units of \(\mathrm{fm}^{n}\) \cite{deVries_1987}. For the 2pF distribution, the parameters are \(c=6.61\,\mathrm{fm}\) and \(a=0.54\,\mathrm{fm}\) \cite{Rdep}; see the text for details. The $n\mathrm{th}$ moments, $\langle r^n \rangle_{c}$ ($n$ = 2, 4, 6, 8, 10, 12), are calculated by using Eq. \eqref{eq:def_moment}. The uncertainty of $\langle r^4 \rangle_{c}$ of the FB distribution is adopted from Ref. \cite{suzuki_2021}.}
     \label{tab:moment_rho}
     {\small
        \begin{tabular}[t]{cccccccc}
           \hline
            Nuclei&$\rho_{c}(r)$&$\langle r^2 \rangle_{c}$&$\langle r^4 \rangle_{c}$&$\langle r^6 \rangle_{c}$&$\langle r^8 \rangle_{c}$&$\langle r^{10} \rangle_{c}$&$\langle r^{12} \rangle_{c}$\\
            \hline
            $^{208}$Pb&FB&30.285&1.172(18)$\times10^3$&5.295$\times10^4$&2.704$\times10^6$&1.542$\times10^8$&9.767$\times10^9$\\
            $^{208}$Pb&SOG&30.285&1.172$\times10^3$&5.294$\times10^4$&2.695$\times10^6$&1.517$\times10^8$&9.331$\times10^9$\\
            
            $^{208}$Pb&2pF&30.234&1.177$\times10^3$&5.423$\times10^4$&2.881$\times10^6$&1.756$\times10^8$&1.233$\times10^{10}$\\
            \hline
    \end{tabular}
    }
    \end{center}
    \end{table}



In order to show how the precision and accuracy of the value of $\langle r^{4}\rangle_{c}$ depends on the above two conditions, we calculate $\left|F_{\mathrm{PWBA}}^{\mathrm{FB}}(q)\right|^{2}$ with the FB distribution \cite{deVries_1987}, according to Eq. \eqref{eq:def_rho}, and use it as the pseudo-experimental form factor, because no experimental data are available in the region of $q < 0.315~\mathrm{fm}^{-1}$. Then, we define the truncated form factor of $\left|F_{\mathrm{PWBA}}^{\mathrm{FB}}(q)\right|^{2}$ including the terms up to $q^{2n}f_{n}$ as $\mathrm{FF}^{n}$ ($n=2,3,4,5,6$):

\allowdisplaybreaks
\begin{subequations}\label{eq:expansion_cut}
    \begin{align}
       \label{eq:expansion_cut_FF2}
       \mathrm{FF^{2}} &= 1 -\frac{\langle r^2 \rangle_c}{3}q^2 + \frac{(3\langle r^4 \rangle_c + 5\langle r^2 \rangle_c^2)}{180}q^4,\\\label{eq:expansion_cut_FF3}
       \mathrm{FF^{3}} &= 1 -\frac{\langle r^2 \rangle_c}{3}q^2 + \frac{(3\langle r^4 \rangle_c + 5\langle r^2 \rangle_c^2)}{180}q^4 -\frac{(\langle r^6 \rangle_c + 7\langle r^4 \rangle_c \langle r^2 \rangle_c)}{2520}q^6,\\
       \mathrm{FF^{4}} &= 1 -\frac{\langle r^2 \rangle_c}{3}q^2 + \frac{(3\langle r^4 \rangle_c + 5\langle r^2 \rangle_c^2)}{180}q^4 -\frac{(\langle r^6 \rangle_c + 7\langle r^4 \rangle_c \langle r^2 \rangle_c)}{2520}q^6 \notag \\
        &\qquad\qquad\qquad\qquad\qquad\quad+ (\frac{\langle r^4 \rangle_c \langle r^4 \rangle_c}{14400} + \frac{\langle r^8 \rangle_c + 12\langle r^2 \rangle_c \langle r^6 \rangle_c}{181440})q^8\\
      \mathrm{FF^{5}} &= 1 -\frac{\langle r^2 \rangle_c}{3}q^2 + \frac{(3\langle r^4 \rangle_c + 5\langle r^2 \rangle_c^2)}{180}q^4 -\frac{(\langle r^6 \rangle_c + 7\langle r^4 \rangle_c \langle r^2 \rangle_c)}{2520}q^6 \notag \\
        &\qquad\qquad\qquad\qquad\qquad\quad+ (\frac{\langle r^4 \rangle_c \langle r^4 \rangle_c}{14400} + \frac{\langle r^8 \rangle_c + 12\langle r^2 \rangle_c \langle r^6 \rangle_c}{181440})q^8 \notag \\
        &\qquad\qquad\qquad\qquad\qquad\quad - (\frac{\langle r^{10} \rangle_c}{19958400}+\frac{\langle r^2 \rangle_c \langle r^8 \rangle_c}{1088640}+\frac{\langle r^4 \rangle_c \langle r^6 \rangle_c}{302400})q^{10},\\
        \mathrm{FF^{6}} &= 1 -\frac{\langle r^2 \rangle_c}{3}q^2 + \frac{(3\langle r^4 \rangle_c + 5\langle r^2 \rangle_c^2)}{180}q^4 -\frac{(\langle r^6 \rangle_c + 7\langle r^4 \rangle_c \langle r^2 \rangle_c)}{2520}q^6 \notag \\
        &\qquad\qquad\qquad\qquad\qquad\quad+ (\frac{\langle r^4 \rangle_c \langle r^4 \rangle_c}{14400} + \frac{\langle r^8 \rangle_c + 12\langle r^2 \rangle_c \langle r^6 \rangle_c}{181440})q^8 \notag \\
        &\qquad\qquad\qquad\qquad\qquad\quad - (\frac{\langle r^{10} \rangle_c}{19958400}+\frac{\langle r^2 \rangle_c \langle r^8 \rangle_c}{1088640}+\frac{\langle r^4 \rangle_c \langle r^6 \rangle_c}{302400})q^{10}\notag \\
        &\qquad\qquad\qquad\qquad\qquad\quad +(\frac{\langle r^{12} \rangle_c}{3113510400} + \frac{\langle r^2 \rangle_c \langle r^{10} \rangle_c}{119750400} + \frac{\langle r^4 \rangle_c \langle r^8 \rangle_c}{21772800} + \frac{\langle r^6 \rangle_c \langle r^6 \rangle_c}{25401600})q^{12}.
    \end{align} 
\end{subequations}
These expressions are rearranged to isolate $\langle r^{4} \rangle_{c}$. For example, Eq. \eqref{eq:expansion_cut_FF3} can be rewritten as 
\begin{align}
\langle r^4\rangle_c
=
\frac{
360\left(\mathrm{FF^{3}}-1+\frac{\langle r^2\rangle_c}{3}q^2-\frac{\langle r^2\rangle_c^2}{36}q^4+\frac{\langle r^6\rangle_c}{2520}q^6\right)
}{
q^4\left(6-\langle r^2\rangle_c q^2\right)
}.
\end{align}
Here, $\mathrm{FF}^{3}$ takes the value of $|F^\mathrm{FB}_\mathrm{PWBA}(q)|^{2}$ as input. We fix $\langle r^{2} \rangle_{c}$ at the value obtained from the FB distribution \cite{deVries_1987}, as listed in Tab. \ref{tab:moment_rho}, and its uncertainty is estimated from the precision given in Ref. \cite{r2_accu}. For the higher-order moment, $\langle r^{6} \rangle_{c}$, we use the value calculated from 
the model-dependent two-parameter Fermi (2pF) density distribution, characterized by the half-density radius $c$ and diffuseness $a$, as listed in Tab.~\ref{tab:moment_rho}.
The reason for employing the 2pF distribution for the higher-order moments is that the moments obtained from the FB and SOG distributions show increasing differences as the order of the moment increases. This is because higher-order moments are sensitive to the tail region of $\rho_{c}(r)$, where the two distributions exhibit different behaviors \cite{3He-3H, 6th}.
As will be shown below, the extracted vale of $\langle r^{4}\rangle_{c}$ is weakly sensitive to whether the higher-order moments of the FB, SOG, or 2pF distribution are used. Therefore, we evaluate the higher-order moments using the 2pF distribution as a simple phenomenological reference.
In this work, $c$ is determined from the empirical $A^{1/3}$ dependence, while a common value of $a$ is assumed for all nuclei~\cite{Rdep}: $c=6.61\,\mathrm{fm}$ and $a=0.54\,\mathrm{fm}$.
For $\mathrm{FF}^{4}$, $\mathrm{FF}^{5}$, and $\mathrm{FF}^{6}$, the same values listed in Tab.~\ref{tab:moment_rho} are also used for $\langle r^{6} \rangle_{c}$ and higher-order moments. 

For each truncated expansion, we extract $\langle r^{4} \rangle_{c}$ at each $q$ value within the convergence region by substituting $|F^\mathrm{FB}_\mathrm{PWBA}(q)|^{2}$ at that $q$ into the truncated expression. 
In this procedure, $\langle r^{2} \rangle_{c}$ is fixed to the value obtained from the FB distribution, whereas the higher-order moments are calculated from the 2pF distribution, as discussed above.
The uncertainty in $\langle r^{4} \rangle_{c}$ is then evaluated by propagating the uncertainties in the scattering cross section, $q$, and $\langle r^{2} \rangle_{c}$, with the cross-section uncertainty propagated to the form factor.
To estimate the uncertainty in $\langle r^{4} \rangle_{c}$ expected in the ongoing measurement, the uncertainties in the scattering cross section and in $q$ are each assumed to be 0.1$\%$. These values correspond to the target experimental precision of the Ultra Low momentum transfer $Q^{2}$ (ULQ$^{2}$) facility at the Research Center for Accelerator and Radioisotope Science (RARiS), Tohoku University \cite{ULQ2, ULQ2_set}.
The uncertainty in $\langle r^{2} \rangle_{c}$ is taken to be 0.05$\%$ \cite{r2_accu}.

\begin{figure*}[!t]
  \centering
    \centering
    \includegraphics[width=0.8\textwidth]{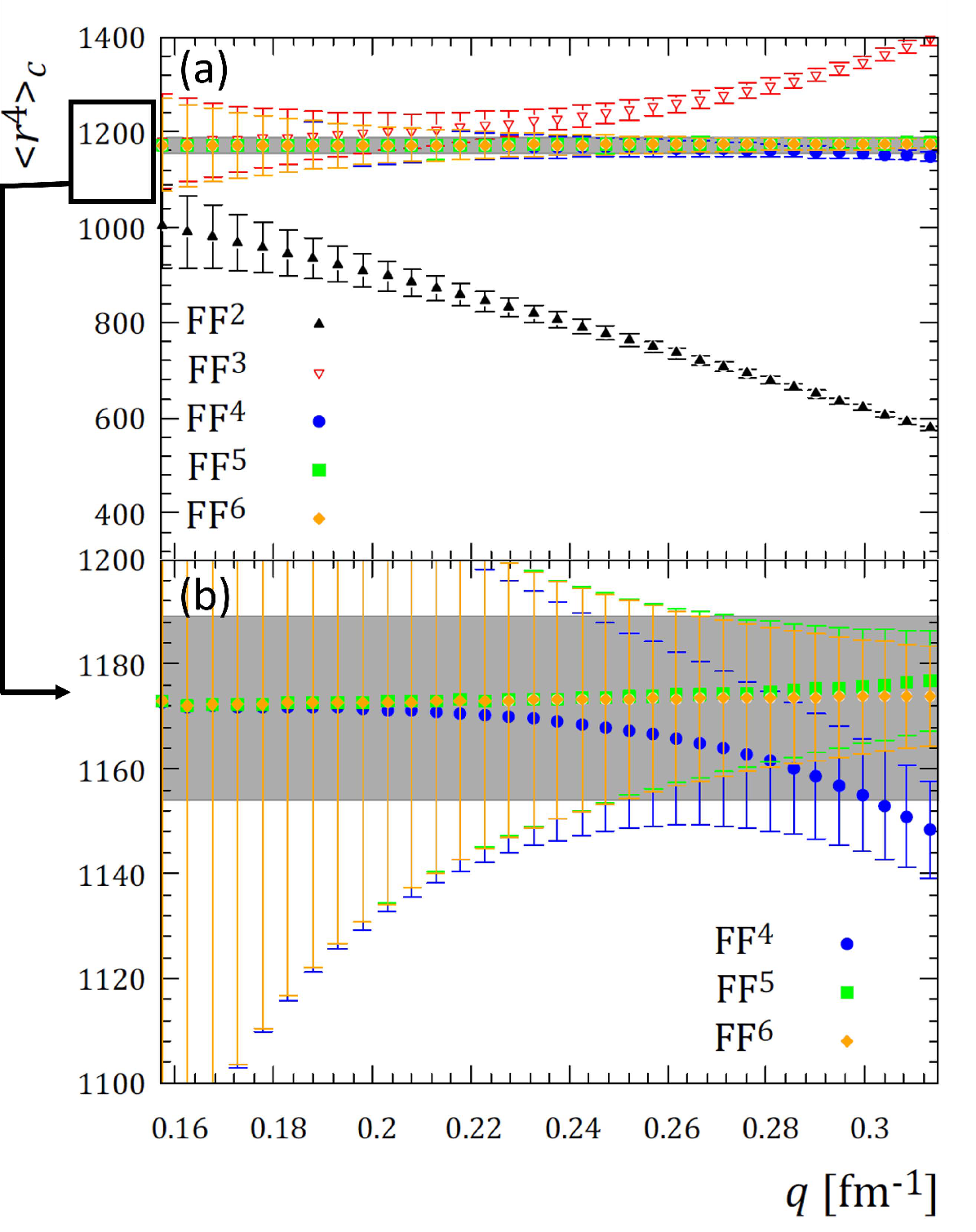}
  \caption{Extracted values of $\langle r^{4}\rangle_{c}$ and their uncertainties obtained with $\mathrm{FF}^{2}$ to $\mathrm{FF}^{6}$. The values obtained with $\mathrm{FF}^{2}$, $\mathrm{FF}^{3}$, $\mathrm{FF}^{4}$, $\mathrm{FF}^{5}$, and $\mathrm{FF}^{6}$ are shown, with each case represented by a different line. The gray band indicates the uncertainty range of  $\langle r^{4}\rangle_{c}$ reported in previous experiments, approximately $1.5\%$ \cite{deVries_1987,suzuki_2021}. Panel (a) shows the results obtained with $\mathrm{FF}^{2}$ to $\mathrm{FF}^{6}$. Panel (b) provides an enlarged view of panel (a), showing only the results obtained with $\mathrm{FF}^{4}$ to $\mathrm{FF}^{6}$.}
  \label{fig:fitting}
\end{figure*}

Figure \ref{fig:fitting} shows the extracted $\langle r^{4} \rangle_{c}$ and their uncertainties at each $q$ value satisfying the convergence condition. As an example, the results for $\mathrm{FF}^{2}$, $\mathrm{FF}^{3}$, $\mathrm{FF}^{4}$, $\mathrm{FF}^{5}$, and $\mathrm{FF}^{6}$ are shown for the case of $E=60\,\mathrm{MeV}$ with $\theta$ varied from $30^\circ$ to $62^\circ$ in steps of $1^\circ$. For comparison, the result of the previous experiment \cite{deVries_1987,suzuki_2021} is also shown as a gray band.
Figure \ref{fig:fitting}(b) shows a vertically enlarged view of Fig.~\ref{fig:fitting}(a), displaying only the results for $\mathrm{FF}^{4}$ to $\mathrm{FF}^{6}$.

For all truncated forms from $\mathrm{FF}^{2}$ to $\mathrm{FF}^{6}$, the precision of \(\langle r^{4} \rangle_{c}\) improves as $q$ increases. To understand the origin of this trend, we consider \(\mathrm{FF}^{2}\) as an example. The uncertainty of $\langle r^{4} \rangle_{c}$ is expressed by

{\footnotesize
\begin{equation}
\label{eq:dr4_FF2}
\delta \langle r^{4} \rangle_{c}
=
\sqrt{
\left(\frac{60}{q^4}\delta\mathrm{FF}^{2}\right)^2
+
\left[
\left(
\frac{20}{q^2}-\frac{10}{3}\langle r^{2} \rangle_{c}
\right)
\delta \langle r^{2} \rangle_{c}
\right]^2
+
\left[
\left(
\frac{40\langle r^{2} \rangle_{c}}{q^3}
-\frac{240}{q^5}
\left(
(\mathrm{FF}^{2}-1)+\frac{\langle r^{2} \rangle_{c}q^2}{3}
\right)
\right)
\delta q
\right]^2.
}
\end{equation}
}
Examining how each term varies with \(q\), the first term, originating from \(\delta\mathrm{FF}^{2}\), decreases as \(1/q^{4}\). The second term, associated with \(\delta\langle r^{2} \rangle_{c}\), is dominated by a $1/q^{2}$ term and therefore becomes less sensitive to $q$ as $q$ increases. The third term, arising from \(\delta q\), varies even more weakly at larger \(q\). Consequently, \(\delta\langle r^{4} \rangle_{c}\) decreases with increasing \(q\) as shown in Fig. ~\ref{fig:fitting}. Thus, from the viewpoint of error propagation, larger \(q\) values are advantageous, and values near the upper limit allowed by the convergence condition are favorable for the precise extraction of \(\langle r^{4} \rangle_{c}\). This provides a basis for selecting the $q$ range used in the analysis.

As shown in Fig.~\ref{fig:fitting}(a), the \(q\) dependence of the extracted \(\langle r^{4} \rangle_{c}\) becomes weaker as higher-order terms are included.
In the case of \(\mathrm{FF}^{2}\), the extracted \(\langle r^{4} \rangle_{c}\) exhibits a clear \(q\) dependence. This indicates that \(\mathrm{FF}^{2}\), namely the expansion up to order \(q^{4}\), is not sufficient to determine \(\langle r^{4} \rangle_{c}\) in the \(q\) range satisfying condition (i).
When \(\mathrm{FF}^{3}\), which includes the \(q^{6}\)-order term, is used, the \(q\) dependence becomes smaller than in the case of \(\mathrm{FF}^{2}\), and it is further suppressed in \(\mathrm{FF}^{4}\).
Moreover, the \(q\) dependence of \(\langle r^{4} \rangle_{c}\) in \(\mathrm{FF}^{3}\) is opposite to that in \(\mathrm{FF}^{2}\): the extracted \(\langle r^{4} \rangle_{c}\) tends to increase with \(q\) in \(\mathrm{FF}^{3}\), whereas it decreases with \(q\) in \(\mathrm{FF}^{2}\). This originates from the added positive contribution of the \(q^{6}\)-order term.

A detailed examination of Fig.~\ref{fig:fitting}(b) shows that the trend observed from $\mathrm{FF}^{2}$ to $\mathrm{FF}^{4}$ also persists from $\mathrm{FF}^{4}$ to $\mathrm{FF}^{6}$.
The comparison of \(\mathrm{FF}^{4}\), \(\mathrm{FF}^{5}\), and \(\mathrm{FF}^{6}\) indicates that the extracted \(\langle r^{4}\rangle_{c}\) has nearly converged at \(\mathrm{FF}^{5}\), yielding a value consistent with the gray band representing the result obtained using the FB distribution in the previous study \cite{deVries_1987, suzuki_2021}; therefore, \(\mathrm{FF}^{5}\) is adopted as the truncated form factor satisfies the condition (ii) for the extraction of \(\langle r^{4} \rangle_{c}\). It should be noted that this agreement indicates that using the higher-order moments, $\langle r^{6}\rangle_{c}$ and above, evaluated from the 2pF distribution does not affect the extracted value of $\langle r^{4}\rangle_{c}$.

These results indicate that, for a precise determination of $\langle r^{4} \rangle_{c}$, the $q$ range should be chosen near the upper limit of the range satisfying the convergence condition, while extraction of $\langle r^{4} \rangle_{c}$ requires the inclusion of terms up to at least order \(q^{10}\), ie, up to \(\mathrm{FF}^{5}\). Under these conditions, measurements around $q=0.3\,\mathrm{fm}^{-1}$ are expected to allow $\langle r^{4} \rangle_{c}$ to be determined with an uncertainty of about 0.8\%, which is substantially better than the currently available uncertainty of about 1.5\% \cite{deVries_1987,suzuki_2021}.

\section{Summary}
We have introduced a new method to determine experimentally the value of \(\langle r^{4}\rangle_{c}\) of \(^{208}\)Pb with improved precision from elastic electron scattering in the low-\(q\) range, where no experimental data are currently available.
By numerically evaluating Coulomb distortion effects and incorporating them into the PWBA framework through a correction factor, \(\left|F_\mathrm{PWBA}(q)\right|^{2}\) can be extracted from the scattering cross section even for a heavy nucleus such as \(^{208}\)Pb. The extracted form factor can then be used to determine \(\langle r^{4}\rangle_{c}\) through its Taylor expansion in the low-$q$ range.

The procedure for determining \(\langle r^{4}\rangle_{c}\) was examined using pseudo form-factor data constructed from the FB charge-density distribution. The analysis shows that an appropriate \(q\) range and truncation order can be selected to achieve high precision with keeping the truncation error negligible. Under the assumed experimental conditions, the uncertainty in the extracted \(\langle r^{4}\rangle_{c}\) is approximately half that obtained in previous studies \cite{deVries_1987, suzuki_2021, ePb}. The resulting uncertainty in \(\sqrt{\langle r^{2}\rangle_{n}}\) is also reduced by approximately 30\% compared with that obtained in previous studies \cite{suzuki_2021}.

An experiment based on this approach is currently in progress at the Ultra Low momentum transfer $Q^{2}$ (ULQ$^{2}$) facility at the Research Center for Accelerator and Radioisotope Science (RARiS), Tohoku University \cite{ULQ2, ULQ2_set}. This measurement will provide an experimental test of the method and is expected to achieve a high-precision determination of $\langle r^{4}\rangle_{c}$. 

Conventional determinations of \(\langle r^{4}\rangle_{c}\) require electron-scattering data over a wide $q$ range, including the high \(q\) region. Since the elastic electron-scattering cross section decreases as \(1/q^{4}\), measurements at high \(q\) require high luminosity. However, achieving such high luminosity is difficult for short-lived unstable nuclei because of the limited number of target nuclei. In this respect, the present method, which makes use of data in the low \(q\) region, may provide a promising approach for future determinations of \(\langle r^{4}\rangle_{c}\) for unstable neutron-rich nuclei through electron scattering at the SCRIT facility at RIKEN \cite{SCRIT}.


\section*{Acknowledgment}
This work was supported by JSPS KAKENHI Grants No. 26K00705, 22K18706, and 20H05635 as
well as the Graduate Program on Physics for the Universe (GP-PU), Tohoku University.


%

\vspace{0.2cm}
\noindent

\let\doi\relax



\end{document}